%% file: main_final.tex
\documentclass[10pt,preprint]{aastex}
\usepackage[utf8]{inputenc}
\usepackage{float}
\usepackage[normalem]{ulem}

\title{Evidence for a systematic offset of $-$0.25~mas in the {\it Gaia\/} DR1 parallaxes}
\author{Keivan G.\ Stassun\altaffilmark{1,2} and Guillermo Torres\altaffilmark{3}}
\altaffiltext{1}{Vanderbilt University, Department of Physics \& Astronomy, 6301 Stevenson Center Ln., Nashville, TN  37235, USA; {\tt keivan.stassun@vanderbilt.edu}}
\altaffiltext{2}{Fisk University, Department of Physics, 1000 17th Ave. N., Nashville, TN  37208, USA}
\altaffiltext{3}{Harvard-Smithsonian Center for Astrophysics, 60 Garden St., Cambridge, MA 02138, USA}

\begin{document}

\begin{abstract}
    We test the parallaxes reported in the {\it Gaia\/} first data release using the sample of eclipsing binaries with accurate, empirical distances from \citet{Stassun:2016}. We find an average offset of $-$0.25$\pm$0.05~mas in the sense of the {\it Gaia\/} parallaxes being too small (i.e., the distances too long). 
    The offset does not depend strongly on obvious parameters such as color or brightness. 
    However, we find with high confidence that the offset may depend on ecliptic latitude: 
    the mean offset is $-$0.38$\pm$0.06~mas in the ecliptic north and $-$0.05$\pm$0.09~mas in the ecliptic south.
    The ecliptic latitude dependence may also be represented by the linear relation, 
    $\Delta\pi \approx -0.22(\pm0.05) -0.003(\pm0.001)\times\beta$~mas ($\beta$ in degrees). 
Finally, there is a possible dependence of the parallax offset on distance, with the offset becoming negligible for $\pi\lesssim 1$~mas; we discuss whether this could be caused by a systematic error in the eclipsing binary distance scale, and reject this interpretation as unlikely. 
\end{abstract}

\section{Introduction\label{sec:intro}}
The advent of trigonometric parallaxes for $\sim$10$^9$ stars from the {\it Gaia\/} mission promises to revolutionize many areas of stellar and Galactic astrophysics, including exoplanet science. 
For example, with eventual expected precision in the parallax of $\approx$20~$\mu$as for bright exoplanet host stars, it should be possible to determine the stellar and planet radii and masses directly and empirically with accuracies of 3--5\% \citep[see, e.g.,][]{Stassun:2016b}. 
Already, the {\it Gaia\/} first data release \citep[DR1;][]{Gaia:2016} provides parallaxes for $\sim$2 million {\it Tycho-2\/} stars (the TGAS stars) with a nominal precision of $\approx$0.3~mas and with a quoted systematic uncertainty at present of 0.3~mas. 

The results from DR1 are based on only 14 months of observations, and use external information in the form of
earlier positions from the {\it Hipparcos\/} \citep{ESA:1997, vanLeeuwen:2007} and {\it Tycho-2\/} \citep{Hog:2000} catalogs to help remove degeneracies \citep[the {\it Tycho-Gaia\/} Astrometric Solution;][]{Michalik:2015}. Additionally, they rely on very provisional and as yet incomplete calibrations, and as a result the astrometric products including the parallaxes are still preliminary. Nonetheless, the new parallaxes represent such an improvement in both quality and quantity that they are certain to be used by the community for a wide range of astrophysical applications, at least until future {\it Gaia\/} releases supersede them.

It is essential, therefore, to assess the on-sky delivered performance of these parallaxes from {\it Gaia\/} DR1, especially the presence of any unexpected biases. This is particularly important in light of the experience from {\it Hipparcos\/}, which suffered a significant bias in at least the case of the Pleiades cluster \citep[e.g.,][]{Pinsonneault:1998}. 
Such a check requires a set of benchmark stars whose parallaxes are determined independently and with an accuracy that is at least as good as that expected from {\it Gaia\/} DR1. 

\citet{Stassun:2016} assembled a sample of 158 eclipsing binary stars (EBs) whose radii and effective temperatures are known empirically and precisely, such that their bolometric luminosities are determined to high precision (via the Stefan-Boltzmann relation) and therefore independent of assumed distance. 
\citet{Stassun:2016} reported new, accurate measurements of the bolometric fluxes for these EBs which, together with the precisely known bolometric luminosities, yields a highly precise distance (or parallax). 
The precision of the parallaxes for this EB sample was predicted by \citet{Stassun:2016} to be $\approx$190~$\mu$as on average. This is a factor of $\sim$1.5 better than the median precision of 320~$\mu$as for {\it Gaia\/} DR1 \citep{Gaia:2016}. It is even somewhat superior to the expected {\it Gaia\/} DR1 precision floor of 240~$\mu$as. 
These EB parallaxes can therefore readily serve as distance benchmarks for the trigonometric parallaxes reported by {\it Gaia} DR1, and in particular can be used to assess the presence of any systematics. 

In this Letter, we report the results of testing the {\it Gaia\/} DR1 parallaxes against the \citet{Stassun:2016} EB benchmark sample. Section~\ref{sec:data} describes the EB and {\it Gaia\/} data used. Section~\ref{sec:results} presents the key result of a systematic offset in the {\it Gaia\/} parallaxes relative to the EB sample. Section~\ref{sec:disc} considers potential trends in the parallax offset with other parameters. Section~\ref{sec:summary} concludes with a summary of our conclusions.

\section{Data\label{sec:data}}
We adopted the predicted parallaxes for the 158 EBs included in the study of \citet{Stassun:2016}. Of these, 116 had parallaxes available in the {\it Gaia\/} first data release.
We excluded from our analysis any EBs identified as potentially problematic in \citet{Stassun:2016}. 
This left 111 EBs with good parallaxes from both the EB analysis and from {\it Gaia}. 
These EBs are all relatively nearby, with parallaxes in the range $\pi \approx$ 0.3--30~mas.
The EBs and their relevant data are provided in Table~\ref{tab:data}.

\begin{deluxetable}{llccrrrrrrrrrrr}
\tablecaption{Eclipsing Binary and {\it Gaia\/} Data\label{tab:data}}
\tablewidth{0pt}
\tablecolumns{16}
\tabletypesize{\scriptsize}
\tablehead{
\colhead{Name} & \colhead{Tycho} & \colhead{$T_{\rm eff}$\tablenotemark{a}} & \colhead{$\sigma_{T_{\rm eff}}$} & \colhead{$V$} & \colhead{$\chi_\nu^2$\tablenotemark{b}} & \colhead{$\pi_{\rm EB}$} & \colhead{$+\sigma_{\pi_{\rm EB}}$} & \colhead{$-\sigma_{\pi_{\rm EB}}$} & \colhead{$\pi_{\rm Gaia}$} & \colhead{$\sigma_{\pi_{\rm Gaia}}$} & \colhead{RA} & \colhead{Dec.} \\
\colhead{} & \colhead{} & \colhead{K} & \colhead{K} & \colhead{mag.} & \colhead{} & \colhead{mas} & \colhead{mas} & \colhead{mas} & \colhead{mas} & \colhead{mas} & \colhead{deg.} & \colhead{deg.}
}
\startdata
UV Psc\tablenotemark{c} & 0026-0577-1 & 5780 & 100 & 9.01 & 2.72 & 12.47 & 0.57 & 0.53 & 14.392 & 0.407 & 19.2297 & 6.8117 \\
XY Cet & 0051-0832-1 & 7870 & 115 & 8.75 & 1.6 & 3.62 & 0.14 & 0.13 & 4.542 & 0.891 & 44.8897 & 3.5176 \\
V1130 Tau & 0066-1108-1 & 6625 & 70 & 6.66 & 1.34 & 14.35 & 0.36 & 0.37 & 14.329 & 0.332 & 57.6748 & 1.5639 \\
EW Ori & 0104-1206-1 & 5970 & 100 & 9.78 & 0.71 & 5.9 & 0.22 & 0.21 & 5.482 & 0.233 & 80.0381 & 2.0444 \\
U Oph\tablenotemark{d} & 0400-1862-1 & 16440 & 250 & 5.9 & 1.15 & 4.35 & 0.14 & 0.17 & 3.685 & 0.775 & 259.1322 & 1.2105 \\
\enddata
\tablecomments{Table \ref{tab:data} is published in its entirety in machine-readable format. A portion is shown here for guidance regarding its form and content.}
\tablenotetext{a}{Effective temperature for the primary component.}
\tablenotetext{b}{$\chi_\nu^2$ of SED fit from \citet{Stassun:2016}. Stars with $\chi_\nu^2 > 20$ were considered unacceptable and are excluded from analysis in this paper also.} 
\tablenotetext{c}{Flagged in \citet{Stassun:2016} as a large outlier relative to {\it Hipparcos\/} and excluded from analysis in this paper.} 
\tablenotetext{d}{Identified in \citet{Stassun:2016} as a known triple system; these are retained in the analysis in this paper (see the text).}
\end{deluxetable}

\section{Results\label{sec:results}}

\begin{figure}[!ht]
\centering
\includegraphics[width=0.7\linewidth,trim=10 0 10 70,clip]{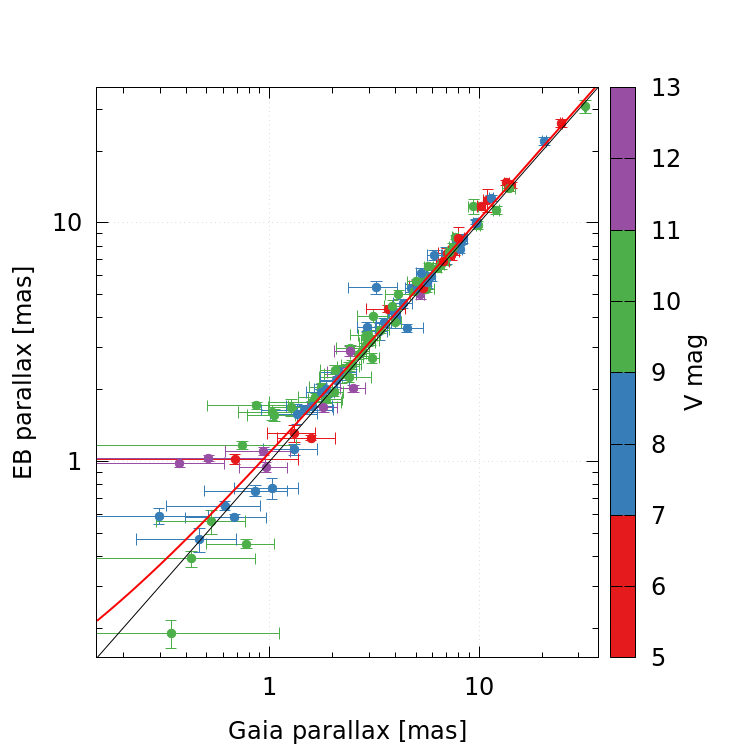}
\caption{Direct comparison of predicted parallaxes from the eclipsing binary sample of \citet{Stassun:2016} versus the parallaxes from the {\it Gaia\/} first data release. The one-to-one line is shown in black and a least-squares linear best fit is shown in red.}
\label{fig:pxvspx} 
\end{figure}

Figure~\ref{fig:pxvspx} shows the direct comparison of the EB parallax predictions from \citet{Stassun:2016} versus the {\it Gaia\/} DR1 parallaxes for the study sample. The least-squares linear best fit, weighted by the measurement uncertainties in both quantities \citep{Press:1992}, is $\pi_{\rm EB} = 0.08 (\pm 0.07) + 1.03 (\pm 0.01) \times \pi_{\rm Gaia}$. While this indicates a good 1-to-1 agreement to first order, the coefficient of 1.03$\pm$0.01 could be interpreted as a modest global difference of {\it scale} in the {\it Gaia\/} parallaxes relative to the EB parallaxes. However, considering all of the available evidence instead suggests a small {\it offset} in the {\it Gaia\/} parallaxes 
as we now discuss. 

\begin{figure}[!ht]
    \centering
    \includegraphics[width=0.7\linewidth,trim=10 10 10 70,clip]{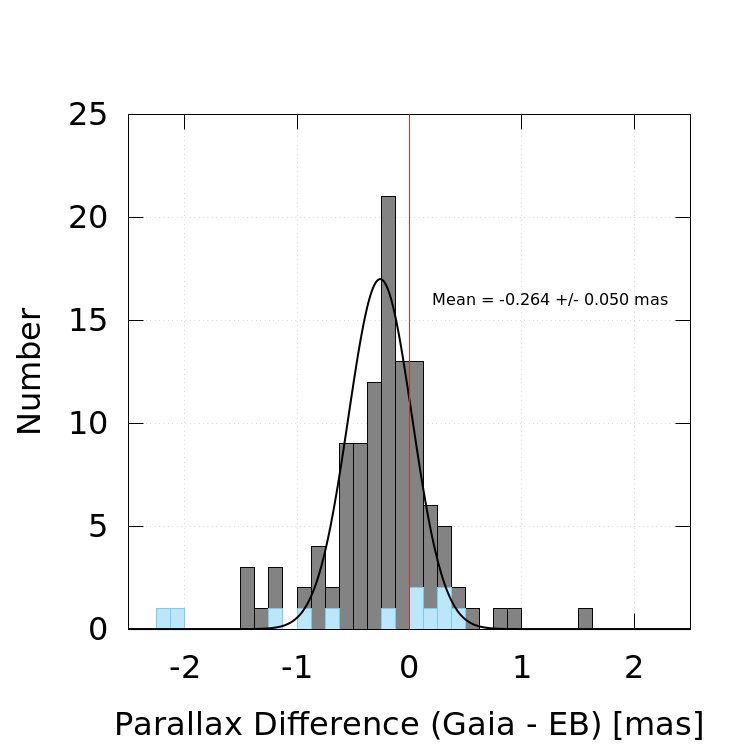}
    \caption{Distribution of $\Delta\pi$ ({\it Gaia\/}$-$EB). EBs with known tertiary companions are represented in blue. We find an offset of $\Delta\pi = -0.264 \pm 0.050$ mas for the entire sample and $\Delta\pi = -0.233 \pm 0.046$ mas when the triple systems are excluded. Also shown is a best-fit Gaussian with $\sigma^2 = \sigma_{\rm Gaia}^2 + \sigma_{\rm EB}^2$, representing the quadrature sum of the typical random uncertainties from the {\it Gaia\/} and EB parallaxes; see Sec.~\ref{sec:intro}.}
    \label{fig:delta_plx_hist}
\end{figure}

Figure~\ref{fig:delta_plx_hist} presents the overall distribution of parallax differences in the sense of $\pi_{\rm Gaia} - \pi_{\rm EB}$. 
The distribution appears roughly symmetric and normally distributed, with perhaps a sharper peak and more extended
wings than a Gaussian, and there is a clear offset relative to zero. 
The mean offset is $\Delta\pi = -0.264 \pm 0.050$ mas, where the quoted error is the uncertainty of the mean for 111 measurements. 

\citet{Stassun:2016} noted that a number of the EBs used in that study are known triple or quadruple systems. In general these companions contribute very little to the total system light, and \citet{Stassun:2016} found no evidence for significant systematics in their predicted parallaxes. Nonetheless, the offset that we find above for the {\it Gaia\/} parallaxes is small and could potentially have gone unnoticed in \citet{Stassun:2016}. Indeed, the effect of additional light contribution to the EBs by companions would be in the sense of making the EB stars appear brighter, therefore inferred to be closer, and in turn the {\it Gaia\/} distances interpreted as too long (parallax too small). 

In {\it Gaia\/} DR1, 12 of our EBs have known companions (see Table~\ref{tab:data}; one additional EB with a known companion is already excluded by the cuts discussed in Sec.~\ref{sec:data}). These EBs are indicated in blue in Fig.~\ref{fig:delta_plx_hist}, which shows the two largest outliers to be triples. Excluding all of the triples results in a parallax offset of 
$\Delta\pi = -0.233 \pm 0.046$ mas, consistent with that determined for the full sample, though slightly smaller. 

Overall, from the EB sample, a systematic offset in the {\it Gaia\/} DR1 parallaxes of $-0.233$ to $-0.264$~mas is indicated. 
For simplicity, we adopt the rounded value between these estimates of $-0.25\pm 0.05$~mas.

\section{Discussion\label{sec:disc}}

The official {\it Gaia\/} DR1 documentation states: 
``There are colour dependent and spatially correlated systematic errors at the level of $\pm$0.2~mas. Over large spatial scales, the parallax zero-point variations reach an amplitude of $\pm$0.3~mas.... Furthermore, averaging parallaxes over small regions of the sky will not reduce the uncertainty on the mean below the 0.3~mas level.\footnote{\url{http://www.cosmos.esa.int/web/gaia/dr1}}"
Our finding of a mean parallax offset of $\Delta\pi\approx -0.25 \pm 0.05$ mas (Sec.~\ref{sec:results}; Fig.~\ref{fig:delta_plx_hist}) corroborates this statement, and further quantifies it using an independent benchmark sample of EBs with accurately known distances \citep{Stassun:2016}. 

In principle this offset could be due to systematics in one or more of the EB parameters from which the EB distances are determined. If so, one might especially suspect the EB $T_{\rm eff}$ values: unlike the stellar radii, for example, which are determined from simple geometry, the $T_{\rm eff}$ values are determined from spectral analysis and/or spectral typing and/or color relations. 
The slope of the fitted relation in Fig.~\ref{fig:pxvspx} would imply an error in the EB distance scale of $\sim$3\%,
which in turn would require a systematic error in $T_{\rm eff}$ of $\sim$1.5\% (because $d \sim L_{\rm bol}^{1/2} \sim T_{\rm eff}^2$) or $\sim$105~K given the typical $T_{\rm eff}$ of the EB sample. 
The sense of the offset is that the EBs would have to be systematically too cool. 

\begin{figure}[!ht]
    \centering
    \includegraphics[width=0.49\linewidth,trim=10 0 10 70,clip]{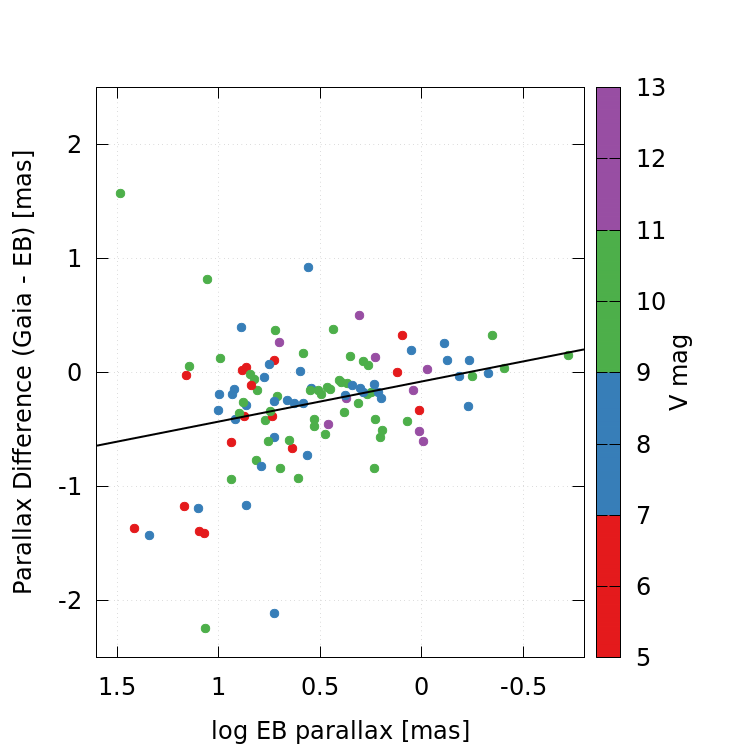}
    \includegraphics[width=0.49\linewidth,trim=10 0 10 70,clip]{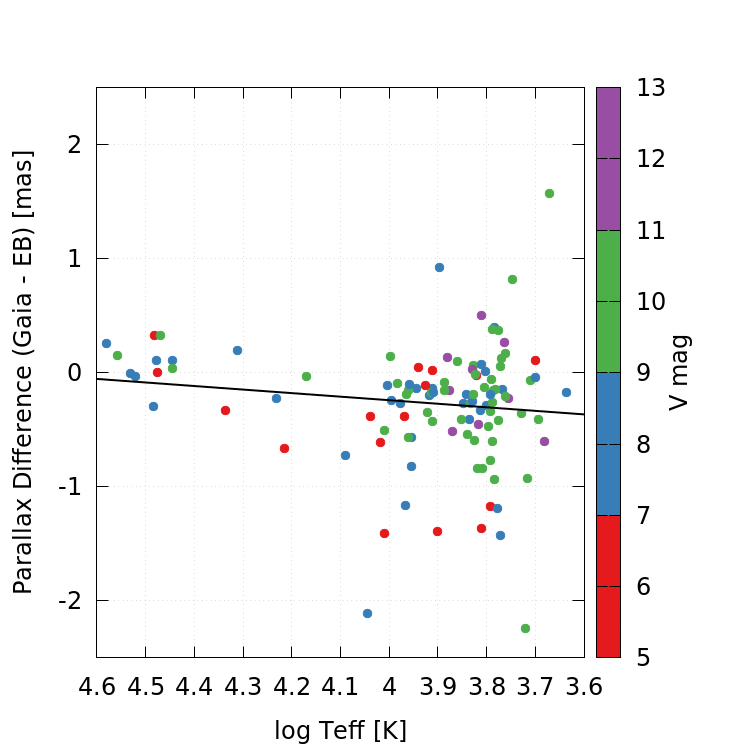}
    \includegraphics[width=0.49\linewidth,trim=10 0 10 70,clip]{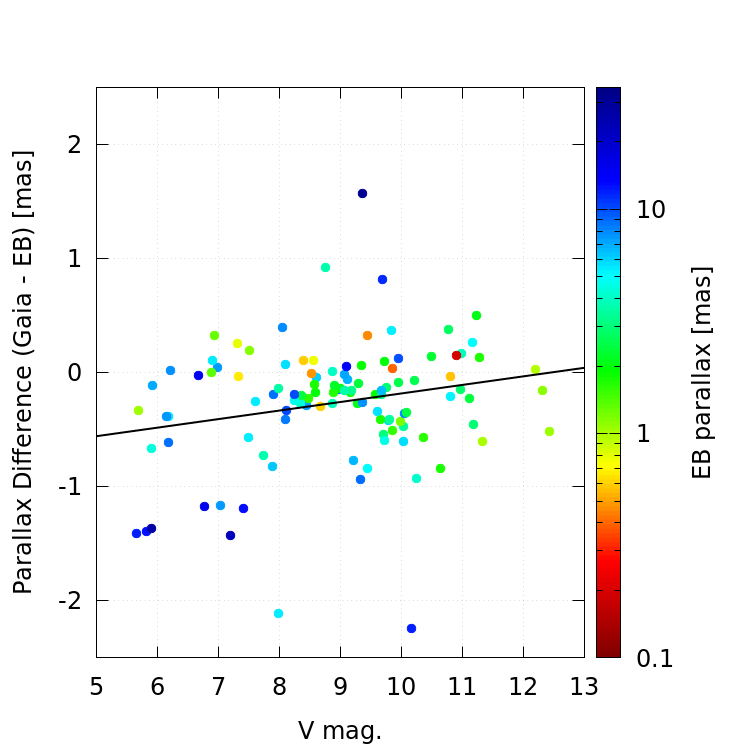}
    \caption{Potential correlations between $\Delta\pi$ and $\pi_{\rm EB}$, EB $\log T_{\rm eff}$, and $V$ mag. Best fitting linear regression lines are shown for each 
    (see the text). }
    \label{fig:delta_vs_eb_plx}
\end{figure}

However, we do not consider this to be a likely possibility, for multiple reasons. 
First, \citet{Stassun:2016} found no evidence for a systematic offset of the EB parallaxes relative to the {\it Hipparcos\/} parallaxes which, even at the somewhat poorer precision of $\sim$1~mas, should have been apparent. 
\citet{Lindegren:2016} compare the {\it Gaia\/} DR1 parallaxes against $\sim$87,000 {\it Hipparcos\/} stars in common, finding a statistically significant average offset just under $-0.1$~mas, smaller than, but in the same sense as the offset we find among our EB sample.
Second, while systematics among various $T_{\rm eff}$ scales can be of order 100~K \citep[see, e.g.,][]{Casagrande:2011,Heiter:2015}, it is unlikely that they should produce a net offset of this entire magnitude in a sample of 111 EBs spanning a large range of $T_{\rm eff}$, given the different methodologies and calibrations adopted by the various authors.
Finally, we have directly examined the degree to which $\Delta\pi$ might correlate with $T_{\rm eff}$ in the EB sample (Fig.~\ref{fig:delta_vs_eb_plx}, upper right), finding very weak evidence for a correlation: 
The regression relation 
has a coefficient of determination $R^2 = 0.13$; this parameter explains only 13\% of the variance in $\Delta\pi$. 
Indeed, a Kendall's $\tau$ non-parametric correlation test gives a probability of 51\% that $\Delta\pi$ and $T_{\rm eff}$ are uncorrelated. (We checked that the parallax {\it ratio} versus $T_{\rm eff}$ is also not significantly correlated.)
Incidentally, this also suggests little to no dependence of $\Delta\pi$ with color, since $T_{\rm eff}$ can be taken as a proxy for color. 

With a current sample of 111 EBs that overlap with the {\it Gaia\/} DR1 parallaxes, it is difficult to ascertain with certainty whether any higher-order dependencies are at work beyond a simple offset. 
%
Nonetheless, it is possible that the parallax offset we observe is 
a function of one or more parameters. We have considered some obvious possible parameters, including distance, and brightness. 
These are represented in Figure~\ref{fig:delta_vs_eb_plx}. 
There is some evidence for a dependence of $\Delta\pi$ on distance and/or brightness, 
however it is difficult to gauge whether these modest distance and brightness dependencies are independent of one another. 
A priori, a possible dependence on brightness may be more likely than a dependence on distance: because of the manner in which the trigonometric parallax measurements are made (linear offsets on the detector), they are more likely to depend on parameters that affect the displacement calibration (e.g., color or brightness) than the displacement itself.
Thus there is not strong evidence for a dependence of the offset with brightness or color, at least in our sample. 


The apparent correlation of $\Delta\pi$ with $\pi$ may be a consequence of the tendency for the EBs with the largest $\pi$ to be located in the ecliptic north, where we find the largest overall offset (see below). In any event, 
among our 12 EBs with $\pi < 1$~mas, the average offset is $\Delta\pi = 0.00 \pm 0.07$~mas.
Thus, it appears that the offset vanishes for very small parallaxes, $\pi \lesssim 1$ mas.
This would be consistent with the findings of \citet{Lindegren:2016} and also \citet{Casertano:2016}, who find good agreement with the {\it Gaia\/} DR1 parallaxes in separate samples of very distant Cepheids with estimates based on period-luminosity relations. \citet{Casertano:2016} also find evidence for a parallax offset, in the same sense as we find, among the small number of very nearby, bright Cepheids in their sample. 

\begin{figure}[!ht]
    \centering
    \includegraphics[width=0.7\linewidth,trim=10 0 10 70,clip]{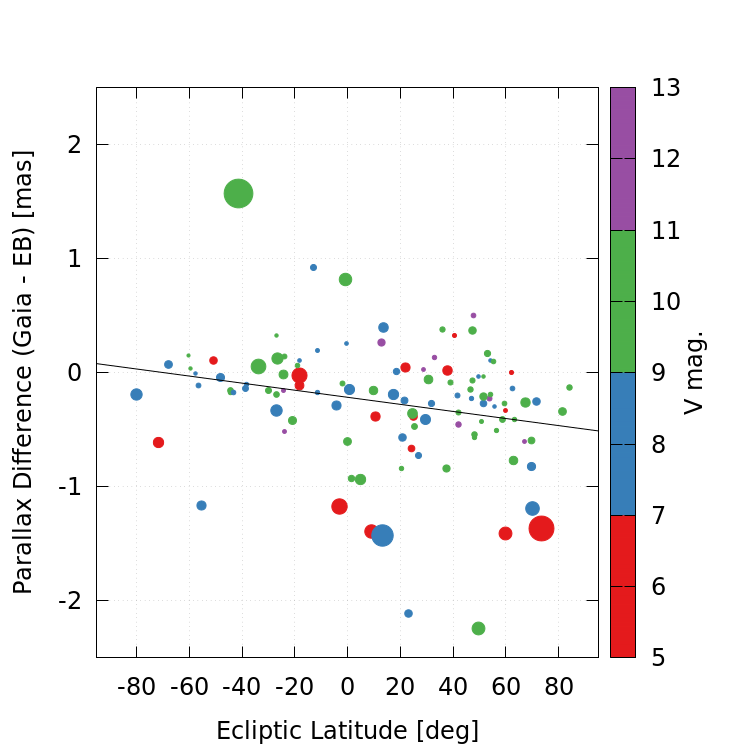}
    \caption{Parallax differences as a function of ecliptic latitude. A trend is found (black line) with 99.7\% confidence, which may also be interpreted as a significant difference in the ecliptic north ($\Delta\pi = -0.38\pm 0.06$ mas) but not in the ecliptic south ($\Delta\pi = -0.05\pm 0.09$ mas). Color represents EB brightness and symbol size is proportional to $\pi$.}
    \label{fig:hist_eclip}
\end{figure}

Finally, we have considered the possible spatial dependence of $\Delta\pi$. 
There is evidence for a trend or difference by ecliptic latitude (Fig.~\ref{fig:hist_eclip}). 
The mean parallax offset for EBs in the northern ecliptic hemisphere is statistically significant with $\Delta\pi = -0.38\pm 0.06$~mas, whereas in the southern ecliptic hemisphere it is not significant with $\Delta\pi = -0.05\pm 0.09$~mas. 
A two-sided Kolmogorov-Smirnov test gives a probability of 0.0001 that this difference could occur by chance. 
Alternatively, the trend can be represented as a linear variation with latitude, $\Delta\pi \approx -0.22(\pm0.05) -0.003(\pm0.001) \times \beta$~mas ($\beta$ in degrees).
A Kendall's $\tau$ test indicates that the correlation between parallax offset and ecliptic latitude is significant with 99.7\% confidence.


The possibility that systematics in the {\it Gaia\/} DR1 parallaxes are dependent on ecliptic latitude was suggested in the {\it Gaia\/} documentation \citep{Lindegren:2016}. These authors reported that a comparison with parallaxes from the {\it Hipparcos\/} mission indicates an overall statistically significant offset $\Delta\pi = -0.089 \pm 0.006$~mas, the {\it Gaia\/} values being smaller, but that the northern ecliptic hemisphere shows a larger offset of $-0.130 \pm 0.006$~mas compared to the southern hemisphere ($-0.053 \pm 0.006$~mas).


\section{Summary and Conclusions\label{sec:summary}}

Here we present evidence of a small but systematic offset in the average
zero-point of the parallax measurements recently released by the
{\it Gaia\/} Mission of about $-0.25 \pm 0.05$~mas, in the sense that
the {\it Gaia\/} values are too small. 
We also find evidence to suggest that the offset is
a function of ecliptic latitude. 
The offset in the northern ecliptic hemisphere is $-$0.38$\pm$0.06~mas and $-$0.05$\pm$0.09~mas in the southern ecliptic hemisphere.
Alternatively, the offset may also be represented as a linear function of the ecliptic latitude,
$\beta$~($^\circ$), according to $\Delta\pi \approx -0.22(\pm0.05) -0.003(\pm0.001)\times\beta$~mas. 

To apply the correction, this (negative) offset must be {\it subtracted} from the reported {\it Gaia\/} DR1 parallaxes. 
At present we can only confirm that the offset is statistically valid for relatively large parallaxes, $\pi \gtrsim 1$~mas.

The reference for this determination is a set of more than 100 independently inferred parallaxes from a benchmark sample of well-studied eclipsing binaries with a wide range of brightnesses and distributed over the entire sky. 
This paper presents evidence of a {\it difference} between the {\it Gaia\/} and EB parallaxes, which we have interpreted here as a systematic error in {\it Gaia\/} after discussing the alternative. In particular, we have considered the possibility of a systematic offset in the EB effective temperature scale 
as a possible, but unlikely alternative explanation. 

It is expected that future releases of the {\it Gaia\/} catalog will remove this small shift as the number of observations increases, calibrations are improved, and the astrometric solution transitions to a self-consistent global fit using only {\it Gaia\/} data, independent of external astrometric information. 
Indeed, these final trigonometric parallaxes may then be used to further refine the EB sample itself, such as improvements to the EB effective temperature scale. 
In the meantime, investigators using the parallax results from {\it Gaia\/} DR1 are encouraged to keep 
the systematic error reported here in mind.

\acknowledgments
This work has made use of the Filtergraph data visualization service at {\tt \url{filtergraph.vanderbilt.edu}} \citep{Burger:2013}. K.G.S.\ acknowledges partial support from NSF PAARE grant AST-1358862. G.T.\ acknowledges partial support for this work from NSF grant AST-1509375.
The authors are grateful to S.~Casertano and A.~Riess for sharing their results in advance of publication. 
We are grateful to the referee for critiques and suggestions that improved the manuscript.
This work has made use of data from the European Space Agency (ESA) mission {\it Gaia\/} (http://www.cosmos.esa.int/gaia), processed by the {\it Gaia\/} Data Processing and Analysis Consortium (DPAC, http://www.cosmos.esa.int/web/gaia/dpac/consortium). Funding for the DPAC has been provided by national institutions, in particular the institutions participating in the {\it Gaia\/} Multilateral Agreement.

\clearpage
\begin{deluxetable}{llccrrrrrrrrrrr}
\tablenum{1}
\tablecaption{Eclipsing Binary and {\it Gaia\/} Data}
\rotate
\tablewidth{0pt}
\tablecolumns{16}
\tabletypesize{\scriptsize}
\tablehead{
\colhead{Name} & \colhead{Tycho} & \colhead{$T_{\rm eff}$\tablenotemark{a}} & \colhead{$\sigma_{T_{\rm eff}}$} & \colhead{$V$} & \colhead{$\chi_\nu^2$\tablenotemark{b}} & \colhead{$\pi_{\rm EB}$} & \colhead{$+\sigma_{\pi_{\rm EB}}$} & \colhead{$-\sigma_{\pi_{\rm EB}}$} & \colhead{$\pi_{\rm Gaia}$} & \colhead{$\sigma_{\pi_{\rm Gaia}}$} & \colhead{RA} & \colhead{Dec.} \\
\colhead{} & \colhead{} & \colhead{K} & \colhead{K} & \colhead{mag.} & \colhead{} & \colhead{mas} & \colhead{mas} & \colhead{mas} & \colhead{mas} & \colhead{mas} & \colhead{deg.} & \colhead{deg.}
}
\startdata
\input gaia_ebs.txt
\enddata
\tablenotetext{a}{Effective temperature for the primary component.}
\tablenotetext{b}{$\chi_\nu^2$ of spectral energy distribution fit from \citet{Stassun:2016}. Stars with $\chi_\nu^2 > 20$ were considered unacceptable and are excluded from analysis in this paper also.} 
\tablenotetext{c}{Flagged in \citet{Stassun:2016} as a large outlier relative to {\it Hipparcos\/} and excluded from analysis in this paper.} 
\tablenotetext{d}{Identified in \citet{Stassun:2016} as a known triple system; these are retained in the analysis in this paper but discussed separately in the text.}
\end{deluxetable}

\end{document}

%% file: gaia_ebs.txt
UV Psc\tablenotemark{c} & 0026-0577-1 & 5780 & 100 & 9.01 & 2.72 & 12.47 & 0.57 & 0.53 & 14.392 & 0.407 & 19.22966316 & 6.8116994 \\
XY Cet & 0051-0832-1 & 7870 & 115 & 8.75 & 1.6 & 3.62 & 0.14 & 0.13 & 4.542 & 0.891 & 44.88971498 & 3.51756508 \\
V1130 Tau & 0066-1108-1 & 6625 & 70 & 6.66 & 1.34 & 14.35 & 0.36 & 0.37 & 14.329 & 0.332 & 57.67475441 & 1.56394837 \\
EW Ori & 0104-1206-1 & 5970 & 100 & 9.78 & 0.71 & 5.9 & 0.22 & 0.21 & 5.482 & 0.233 & 80.0381137 & 2.0444353 \\
V578 Mon & 0154-2528-1 & 30000 & 740 & 8.55 & 0.65 & 0.75 & 0.04 & 0.04 & 0.855 & 0.364 & 98.0025408 & 4.8780283 \\
AI Hya & 0196-0626-1 & 6700 & 60 & 9.35 & 7.02 & 1.82 & 0.07 & 0.07 & 1.879 & 0.353 & 124.697754 & 0.283378 \\
FM Leo & 0263-0727-1 & 6316 & 240 & 8.45 & 3.49 & 7.29 & 0.64 & 0.57 & 7.004 & 0.322 & 168.18789591 & 0.34800926 \\
AQ Ser & 0340-0588-1 & 6430 & 100 & 10.65 & 1.89 & 1.71 & 0.06 & 0.06 & 0.869 & 0.366 & 230.5632692 & 2.50308 \\
V335 Ser & 0353-0301-1 & 9020 & 150 & 7.49 & 1.99 & 5.31 & 0.2 & 0.21 & 4.742 & 0.302 & 239.7739821 & 0.5957072 \\
U Oph\tablenotemark{d} & 0400-1862-1 & 16440 & 250 & 5.9 & 1.15 & 4.35 & 0.14 & 0.17 & 3.685 & 0.775 & 259.13215193 & 1.21054491 \\
V2368 Oph & 0404-2156-1 & 9300 & 200 & 6.18 & 2.84 & 5.44 & 0.32 & 0.29 & 5.06 & 0.428 & 259.05941952 & 2.18620241 \\
V413 Ser\tablenotemark{d} & 0446-0091-1 & 11100 & 300 & 7.99 & 1.64 & 5.34 & 0.35 & 0.32 & 3.23 & 0.853 & 278.78422403 & 0.04301045 \\
CoRoT 105906206 & 0459-0892-1 & 6750 & 150 & 12.21 & 1.8 & 0.94 & 0.05 & 0.04 & 0.964 & 0.249 & 280.820621 & 5.966736 \\
IO Aqr\tablenotemark{d} & 0511-0960-1 & 6336 & 125 & 8.86 & 0.87 & 3.97 & 0.18 & 0.17 & 3.979 & 0.246 & 310.18946139 & 0.93917186 \\
V1388 Ori & 0738-0244-1 & 20500 & 500 & 7.5 & 3.28 & 1.12 & 0.06 & 0.06 & 1.315 & 0.381 & 92.74652296 & 11.99485667 \\
WZ Oph & 0977-0216-1 & 6165 & 100 & 9.12 & 3.6 & 6.67 & 0.27 & 0.27 & 6.608 & 0.238 & 256.66267585 & 7.78271678 \\
V2365 Oph & 0977-0547-1 & 9500 & 200 & 8.86 & 1.68 & 3.81 & 0.19 & 0.16 & 3.542 & 0.29 & 257.19075987 & 9.1861524 \\
V624 Her & 1005-2131-1 & 8150 & 150 & 6.2 & 2.11 & 7.6 & 0.29 & 0.35 & 7.615 & 0.489 & 266.07186051 & 14.41006621 \\
EE Peg\tablenotemark{d} & 1120-0161-1 & 8700 & 200 & 6.98 & 1.72 & 7.33 & 0.41 & 0.32 & 7.372 & 0.298 & 325.00784071 & 9.18475409 \\
CF Tau\tablenotemark{d} & 1262-0050-1 & 5200 & 150 & 10.24 & 1.31 & 4.06 & 0.26 & 0.25 & 3.132 & 0.513 & 61.2922271 & 22.4967133 \\
V1094 Tau & 1263-0642-1 & 5850 & 100 & 8.97 & 0.74 & 8.4 & 0.34 & 0.3 & 8.256 & 0.251 & 63.0149692 & 21.9473753 \\
CD Tau\tablenotemark{d} & 1291-0292-1 & 6200 & 50 & 6.77 & 4.36 & 14.73 & 0.41 & 0.39 & 13.559 & 0.375 & 79.379796 & 20.131842 \\
FT Ori & 1326-0910-1 & 9600 & 400 & 9.29 & 7.62 & 2.32 & 0.22 & 0.19 & 2.222 & 0.383 & 93.4923067 & 21.4275506 \\
BP Vul & 1644-2113-1 & 7715 & 150 & 9.95 & 2.61 & 2.49 & 0.11 & 0.1 & 2.402 & 0.294 & 306.38853349 & 21.0383264 \\
AD Boo & 2015-0216-1 & 6575 & 120 & 9.44 & 2.31 & 4.99 & 0.2 & 0.19 & 4.146 & 0.554 & 218.8032496 & 24.6392644 \\
RT CrB & 2039-1337-1 & 5134 & 100 & 10.22 & 1.84 & 2.54 & 0.11 & 0.11 & 2.475 & 0.268 & 234.5126246 & 29.48720441 \\
LV Her & 2076-1042-1 & 6060 & 150 & 10.97 & 1.17 & 2.82 & 0.16 & 0.15 & 2.675 & 0.243 & 263.8850083 & 23.1751667 \\
DI Her & 2109-0775-1 & 17000 & 800 & 8.47 & 0.65 & 1.58 & 0.16 & 0.15 & 1.354 & 0.346 & 283.35933236 & 24.27799733 \\
BK Peg & 2254-2563-1 & 6265 & 85 & 10.04 & 0.82 & 3.38 & 0.11 & 0.1 & 2.911 & 0.482 & 356.7852704 & 26.5666433 \\
AR Aur & 2398-1311-1 & 10950 & 300 & 6.14 & 1.83 & 7.43 & 0.43 & 0.41 & 7.047 & 0.611 & 79.57875162 & 33.76734788 \\
HP Aur\tablenotemark{d} & 2401-1263-1 & 5810 & 120 & 11.17 & 1.31 & 5.01 & 0.23 & 0.21 & 5.273 & 0.22 & 77.590762 & 35.796286 \\
V432 Aur & 2416-0768-1 & 6080 & 85 & 8.05 & 4.61 & 7.72 & 0.29 & 0.28 & 8.118 & 0.255 & 84.38545384 & 37.08673876 \\
WW Aur & 2426-0345-1 & 7960 & 420 & 5.82 & 3.06 & 12.42 & 1.46 & 1.26 & 11.026 & 0.5 & 98.11326959 & 32.45489819 \\
KX Cnc & 2484-0592-1 & 5900 & 100 & 7.19 & 1.05 & 21.97 & 0.83 & 0.77 & 20.542 & 0.378 & 130.69254478 & 31.86260197 \\
HD 71636 & 2489-1972-1 & 6950 & 140 & 7.9 & 0.92 & 8.59 & 0.41 & 0.35 & 8.403 & 0.402 & 127.48463115 & 37.07096799 \\
CV Boo & 2570-0843-1 & 5760 & 150 & 10.99 & 1.64 & 3.83 & 0.23 & 0.21 & 3.997 & 0.253 & 231.5813963 & 36.9815092 \\
V501 Her & 2606-1905-1 & 5683 & 100 & 11.12 & 0.5 & 2.36 & 0.09 & 0.09 & 2.129 & 0.231 & 263.93108 & 30.64308 \\
V885 Cyg & 2655-1877-1 & 8150 & 150 & 9.99 & 2.27 & 1.17 & 0.05 & 0.04 & 0.744 & 0.608 & 293.2077342 & 30.0213972 \\
MY Cyg & 2680-1529-1 & 7050 & 200 & 8.34 & 1.79 & 4.22 & 0.27 & 0.25 & 3.952 & 0.235 & 305.01412547 & 33.94306105 \\
V453 Cyg & 2683-3326-1 & 27800 & 400 & 8.4 & 1.62 & 0.58 & 0.02 & 0.02 & 0.683 & 0.288 & 301.6456979 & 35.7406322 \\
V442 Cyg & 2685-1903-1 & 6900 & 100 & 9.7 & 1.68 & 2.98 & 0.1 & 0.09 & 2.44 & 0.346 & 306.9679333 & 30.791195 \\
CG Cyg\tablenotemark{d} & 2696-2945-1 & 5260 & 180 & 10.16 & 3.26 & 11.68 & 0.89 & 0.8 & 9.438 & 0.545 & 314.556071 & 35.174911 \\
Y Cyg & 2696-3486-1 & 33200 & 600 & 7.32 & 1.58 & 0.65 & 0.03 & 0.03 & 0.613 & 0.291 & 313.01490655 & 34.65763334 \\
KIC 8410637 & 3130-2385-1 & 4800 & 80 & 11.33 & 2.63 & 0.98 & 0.04 & 0.04 & 0.373 & 0.238 & 282.1587429 & 44.4860661 \\
KIC 3858884 & 3135-0651-1 & 6800 & 70 & 9.28 & 10.6 & 2.05 & 0.07 & 0.07 & 1.776 & 0.224 & 293.69543217 & 38.98278969 \\
V380 Cyg & 3141-3692-1 & 21700 & 400 & 5.68 & 0.9 & 1.02 & 0.05 & 0.04 & 0.685 & 0.696 & 297.65553032 & 40.59976051 \\
V478 Cyg & 3151-2222-1 & 30479 & 1000 & 8.68 & 5.82 & 0.59 & 0.05 & 0.04 & 0.297 & 0.214 & 304.91144903 & 38.33588957 \\
V364 Lac & 3215-0971-1 & 8250 & 150 & 8.36 & 2.3 & 2.37 & 0.1 & 0.1 & 2.166 & 0.423 & 343.06170498 & 38.74573119 \\
V342 And\tablenotemark{d} & 3246-2531-1 & 6200 & 100 & 7.82 & 24.15 & 12.15 & 0.69 & 0.69 & 4.915 & 0.333 & 2.5133037 & 46.3903136 \\
CO And\tablenotemark{d} & 3268-0398-1 & 6140 & 130 & 10.77 & 1.34 & 2.71 & 0.13 & 0.13 & 3.087 & 0.274 & 17.8534604 & 46.9637042 \\
V570 Per & 3314-1225-1 & 6842 & 50 & 8.09 & 1.98 & 8.26 & 0.19 & 0.18 & 7.847 & 0.264 & 47.39559997 & 48.62463695 \\
IM Per\tablenotemark{d} & 3323-1123-1 & 7580 & 150 & 11.28 & 3.91 & 1.68 & 0.09 & 0.07 & 1.814 & 0.291 & 47.9262975 & 52.2117222 \\
IQ Per & 3331-1175-1 & 12300 & 230 & 7.73 & 3.2 & 3.65 & 0.16 & 0.15 & 2.92 & 0.281 & 59.93615716 & 48.15124737 \\
HS Aur & 3394-0326-1 & 5350 & 75 & 10.05 & 1.71 & 7.89 & 0.3 & 0.29 & 7.53 & 0.282 & 102.82698688 & 47.67338091 \\
FL Lyr & 3542-1492-1 & 6150 & 100 & 9.36 & 1.53 & 7.51 & 0.33 & 0.29 & 7.25 & 0.217 & 288.02025709 & 46.32412867 \\
V2080 Cyg & 3551-1744-1 & 6000 & 75 & 7.4 & 2.15 & 12.63 & 0.36 & 0.35 & 11.439 & 0.252 & 291.69977628 & 50.14549388 \\
KIC 9246715 & 3559-0102-1 & 4930 & 190 & 9.65 & 3.92 & 1.68 & 0.15 & 0.13 & 1.272 & 0.279 & 300.9513421 & 45.6041192 \\
V1061 Cyg & 3600-0472-1 & 6180 & 100 & 9.21 & 2.29 & 6.54 & 0.26 & 0.21 & 5.768 & 0.291 & 316.83549669 & 52.04956061 \\
RW Lac & 3629-0740-1 & 5760 & 100 & 10.81 & 1.83 & 5.15 & 0.21 & 0.2 & 4.942 & 0.301 & 341.2378975 & 49.6576586 \\
AP And & 3639-0915-1 & 6565 & 150 & 11.19 & 1.02 & 2.89 & 0.15 & 0.14 & 2.433 & 0.388 & 357.3779521 & 45.7892364 \\
IT Cas & 3650-0959-1 & 6470 & 100 & 11.23 & 1.13 & 2.02 & 0.07 & 0.07 & 2.52 & 0.339 & 355.5058137 & 51.7435558 \\
V505 Per\tablenotemark{c} & 3690-0536-1 & 6510 & 50 & 6.88 & 0.99 & 16.93 & 0.41 & 0.37 & 15.563 & 0.323 & 35.30401536 & 54.51007761 \\
V1143 Cyg & 3938-1983-1 & 6450 & 100 & 5.9 & 2.02 & 26.11 & 0.96 & 0.88 & 24.746 & 0.354 & 294.6715988 & 54.97379118 \\
RT And\tablenotemark{c} & 3998-2167-1 & 6100 & 150 & 9.04 & 5.42 & 9.17 & 0.54 & 0.5 & 10.053 & 0.225 & 347.79207917 & 53.02584476 \\
V396 Cas & 4006-1219-1 & 9225 & 150 & 9.58 & 1.82 & 1.85 & 0.08 & 0.06 & 1.661 & 0.279 & 348.3999133 & 56.7381106 \\
PV Cas & 4010-1411-1 & 10200 & 250 & 9.86 & 3.77 & 1.56 & 0.09 & 0.08 & 1.057 & 0.276 & 347.51073 & 59.2017072 \\
MU Cas & 4014-1119-1 & 14750 & 800 & 10.8 & 3.1 & 0.56 & 0.07 & 0.06 & 0.528 & 0.24 & 3.96483911 & 60.43156742 \\
V459 Cas & 4030-1001-1 & 9140 & 300 & 10.36 & 3.94 & 1.6 & 0.12 & 0.11 & 1.028 & 0.318 & 17.8746546 & 61.1466544 \\
SZ Cam\tablenotemark{d} & 4068-1651-1 & 30320 & 150 & 6.93 & 3.57 & 1.25 & 0.03 & 0.03 & 1.578 & 0.484 & 61.95538446 & 62.33293757 \\
WW Cam & 4073-1191-1 & 8350 & 135 & 10.09 & 3.26 & 2.41 & 0.11 & 0.1 & 2.062 & 0.303 & 67.8553362 & 64.3626378 \\
ZZ UMa & 4144-0400-1 & 5960 & 70 & 9.83 & 1.74 & 5.27 & 0.15 & 0.14 & 5.638 & 0.469 & 157.5133057 & 61.81150639 \\
WX Cep & 4268-0138-1 & 8150 & 250 & 9 & 12.47 & 2.01 & 0.16 & 0.14 & 1.873 & 0.234 & 337.81577827 & 63.52265463 \\
AH Cep & 4273-0857-1 & 29900 & 1000 & 6.88 & 5.84 & 1.31 & 0.11 & 0.1 & 1.315 & 0.341 & 341.97059463 & 65.06216647 \\
YZ Cas & 4307-2167-1 & 10200 & 300 & 5.65 & 4.63 & 11.71 & 0.41 & 0.37 & 10.3 & 0.488 & 11.41282179 & 74.98807249 \\
BF Dra & 4435-1750-1 & 6360 & 150 & 9.76 & 1.42 & 2.9 & 0.16 & 0.13 & 2.77 & 0.224 & 282.74730236 & 69.88263435 \\
UZ Dra & 4444-1595-1 & 6200 & 100 & 9.6 & 4.95 & 5.55 & 0.24 & 0.23 & 5.206 & 0.253 & 291.4793554 & 68.9353192 \\
EK Cep & 4466-2120-1 & 9000 & 200 & 7.89 & 4.61 & 6.14 & 0.32 & 0.3 & 5.316 & 0.308 & 325.33960392 & 69.69280812 \\
VZ Cep & 4470-1334-1 & 6670 & 160 & 9.72 & 1.82 & 4.47 & 0.25 & 0.24 & 3.876 & 0.348 & 327.5463979 & 71.4439708 \\
EY Cep & 4521-0349-1 & 7090 & 150 & 9.81 & 0.77 & 3.37 & 0.16 & 0.15 & 2.957 & 0.314 & 55.0169708 & 81.0191858 \\
AY Cam & 4540-1742-1 & 7250 & 100 & 9.72 & 0.99 & 1.94 & 0.06 & 0.06 & 2.035 & 0.225 & 126.4657754 & 77.2185686 \\
EI Cep & 4599-0082-1 & 6750 & 100 & 7.61 & 1.93 & 5.32 & 0.2 & 0.17 & 5.066 & 0.243 & 322.11752646 & 76.40349679 \\
GG Ori & 4767-0857-1 & 9950 & 200 & 10.49 & 3.43 & 2.25 & 0.11 & 0.1 & 2.396 & 0.673 & 85.792592 & $-$0.687461 \\
V530 Ori & 4786-0571-1 & 5890 & 100 & 9.96 & 1.09 & 9.77 & 0.41 & 0.38 & 9.897 & 0.224 & 91.1408646 & $-$3.1976719 \\
V501 Mon & 4799-1943-1 & 7510 & 100 & 12.32 & 2.48 & 1.1 & 0.04 & 0.04 & 0.939 & 0.325 & 100.173871 & $-$1.111114 \\
CoRoT 102918586 & 4800-1540-1 & 7400 & 90 & 12.43 & 0.99 & 1.03 & 0.03 & 0.03 & 0.513 & 0.443 & 102.226296 & $-$0.873125 \\
HI Mon & 4809-0245-1 & 29500 & 600 & 9.45 & 4.1 & 0.45 & 0.02 & 0.02 & 0.776 & 0.274 & 103.9544438 & $-$4.0432744 \\
FS Mon & 4825-2374-1 & 6715 & 100 & 9.68 & 2.17 & 3.13 & 0.1 & 0.1 & 2.937 & 0.261 & 111.1762621 & $-$5.1540478 \\
VZ Hya & 4874-0811-1 & 6645 & 150 & 9.06 & 0.8 & 6.95 & 0.35 & 0.32 & 6.936 & 0.235 & 127.9225553 & $-$6.31876784 \\
IM Vir & 4955-0912-1 & 5570 & 100 & 9.69 & 1.39 & 11.3 & 0.48 & 0.45 & 12.12 & 0.342 & 192.4112337 & $-$6.0791283 \\
BH Vir & 4968-0569-1 & 6100 & 100 & 9.68 & 1.81 & 6.46 & 0.28 & 0.27 & 6.306 & 0.296 & 209.60358735 & $-$1.66082075 \\
EG Ser & 5099-0149-1 & 9900 & 200 & 8.24 & 4.57 & 4.59 & 0.21 & 0.2 & 4.342 & 0.445 & 276.5091796 & $-$1.6809489 \\
LL Aqr & 5236-0883-1 & 6080 & 45 & 9.32 & 0.9 & 8.68 & 0.19 & 0.19 & 7.746 & 0.271 & 338.67563291 & $-$3.5994906 \\
EF Aqr & 5248-1030-1 & 6150 & 65 & 10.04 & 0.67 & 5.66 & 0.15 & 0.14 & 5.056 & 0.499 & 345.3295333 & $-$6.4375969 \\
PV Pup & 5422-3294-1 & 6920 & 300 & 6.93 & 85.13 & 11.98 & 1.49 & 1.28 & 11.969 & 0.334 & 116.36971411 & $-$14.68613454 \\
GZ CMa & 5965-0860-1 & 8800 & 350 & 7.98 & 2.25 & 3.5 & 0.32 & 0.28 & 3.36 & 0.311 & 109.08002714 & $-$16.71669043 \\
SW CMa & 5976-0630-1 & 8200 & 150 & 9.16 & 3.87 & 1.77 & 0.07 & 0.08 & 1.596 & 0.601 & 107.06348559 & $-$22.4403521 \\
HW CMa & 5976-1266-1 & 7700 & 150 & 9.18 & 2.14 & 3.21 & 0.15 & 0.14 & 3.056 & 0.305 & 107.0910854 & $-$22.4082975 \\
HS Hya & 6069-1131-1 & 6500 & 50 & 8.12 & 1.04 & 10.01 & 0.23 & 0.16 & 9.676 & 0.268 & 156.15319885 & $-$19.09248835 \\
AK For & 6446-0342-1 & 4690 & 100 & 9.36 & 5.68 & 30.65 & 1.88 & 1.8 & 32.222 & 0.247 & 52.34530926 & $-$24.10085907 \\
V3903 Sgr & 6843-0543-1 & 38000 & 1900 & 7.31 & 0.47 & 0.77 & 0.08 & 0.07 & 1.029 & 0.348 & 272.32374524 & $-$23.98839603 \\
TZ For & 7026-0633-1 & 5000 & 100 & 6.89 & 0.75 & 5.33 & 0.25 & 0.21 & 5.436 & 0.254 & 48.66705279 & $-$35.55766586 \\
HD 187669 & 7443-0867-1 & 4330 & 70 & 8.88 & 1.13 & 1.64 & 0.07 & 0.07 & 1.467 & 0.549 & 298.0920162 & $-$32.5610378 \\
PT Vel & 7690-2859-1 & 9250 & 150 & 7.03 & 13.27 & 7.31 & 0.38 & 0.35 & 6.145 & 0.451 & 137.74049908 & $-$43.26748036 \\
V4089 Sgr & 7936-2270-1 & 8433 & 97 & 5.91 & 2.66 & 6.86 & 0.18 & 0.18 & 6.748 & 0.489 & 293.53535674 & $-$40.03463968 \\
AI Phe & 8032-0625-1 & 5010 & 120 & 8.6 & 0.52 & 5.98 & 0.31 & 0.3 & 5.938 & 0.238 & 17.39247871 & $-$46.26558121 \\
V467 Vel & 8151-1072-1 & 36200 & 2500 & 10.9 & 10.29 & 0.19 & 0.03 & 0.02 & 0.339 & 0.777 & 130.95325 & $-$46.125928 \\
V636 Cen & 8285-0847-1 & 5900 & 85 & 9.09 & 0.25 & 13.91 & 0.43 & 0.4 & 13.962 & 0.991 & 214.24130865 & $-$49.94510016 \\
TV Nor & 8322-0334-1 & 9120 & 150 & 9.06 & 1.99 & 3.55 & 0.13 & 0.12 & 3.389 & 0.33 & 241.0385217 & $-$51.5444425 \\
GV Car & 8627-1797-1 & 10100 & 300 & 8.91 & 2.72 & 2.19 & 0.17 & 0.13 & 2.075 & 0.325 & 166.387038 & $-$58.730517 \\
SZ Cen & 8676-2330-1 & 8100 & 300 & 8.59 & 17.76 & 1.94 & 0.18 & 0.16 & 1.769 & 0.263 & 207.646219 & $-$58.49919382 \\
DW Car & 8957-1314-1 & 27900 & 1000 & 9.85 & 3.76 & 0.39 & 0.03 & 0.03 & 0.423 & 0.432 & 160.79195 & $-$60.0365947 \\
EM Car & 8959-0569-1 & 34000 & 2000 & 8.52 & 0.99 & 0.47 & 0.06 & 0.05 & 0.464 & 0.232 & 168.0187762 & $-$61.0952586 \\
V349 Ara & 9038-0641-1 & 9074 & 200 & 8.58 & 3.91 & 1.7 & 0.09 & 0.09 & 1.6 & 0.392 & 249.8442192 & $-$60.9617075 \\
UX Men\tablenotemark{d} & 9378-0190-1 & 6200 & 100 & 8.24 & 1.15 & 9.91 & 0.39 & 0.38 & 9.718 & 0.214 & 82.51326739 & $-$76.24870878 \\
RZ Cha & 9422-0104-1 & 6450 & 150 & 8.09 & 0.68 & 5.61 & 0.28 & 0.28 & 5.685 & 0.259 & 160.60043486 & $-$82.0372742 \\
TZ Men & 9496-0590-1 & 10400 & 500 & 6.18 & 2.04 & 8.63 & 1.01 & 0.81 & 8.022 & 0.488 & 82.557857 & $-$84.78510291 \\